\documentclass[aps,pra,reprint,superscriptaddress,showpacs]{revtex4-1}

\usepackage{graphicx}
\usepackage{amsmath}
\usepackage{bm}
\usepackage{graphicx}
\usepackage{dcolumn}
\usepackage{bm}% bold math
\usepackage{colordvi}
\usepackage{color}
\usepackage{ulem}
\usepackage{type1cm}
 \usepackage{lettrine}
\newcommand{\beq}{\begin{equation}}
\newcommand{\eeq}{\end{equation}}
\newcommand{\beqa}{\begin{eqnarray}}
\newcommand{\eeqa}{\end{eqnarray}}
\def\ra{\rangle}

\begin{document}

%\title{SUBJECT AREA: Quantum dots, Spin control, Shortcuts to adiabaticity}

\title{Counter-diabatic driving for fast spin control in a two-electron double quantum dot}

\author{Yue Ban}
\email{yban@shu.edu.cn}
\affiliation{Department of Electronic Information Materials, Shanghai University, 200444 Shanghai, People's Republic of China }

\author{Xi Chen}
\email{xchen@shu.edu.cn}
\affiliation{Department of Physics, Shanghai University, 200444 Shanghai, People's Republic of China }

\date{\today}
\begin{abstract}
The techniques of shortcuts to adiabaticity have been proposed to accelerate the ``slow" adiabatic processes in various quantum systems
with the applications in quantum information processing.
In this paper, we study the counter-diabatic driving for fast adiabatic spin manipulation in a two-electron double quantum dot
by designing time-dependent electric fields in the presence of spin-orbit coupling. To simplify implementation and find an alternative shortcut,
we further transform the Hamiltonian in term of Lie algebra, which allows one to use a single Cartesian component of electric fields.
In addition, the relation between energy and time is quantified to show the lower bound for the operation time when the maximum amplitude of electric fields is given.
Finally, the fidelity is discussed with respect to noise and systematic errors, which demonstrates that the decoherence effect induced by stochastic environment
can be avoided in speeded-up adiabatic control.

\end{abstract}
%
%\pacs{03.67.Ac, 73.63.Kv}%

\maketitle

\section*{Introduction}
\lettrine{E}{lectron} spins in quantum dots (QDs) \cite{spin resonance,Rashba,Nowack,You-Sun,Guo} have been extensively investigated for potential applications in quantum information processing, as spins in QDs are expected as a possible realization of qubit in quantum information science and technology \cite{QD-Qgates}. Especially, a two-electron double QD can be further regarded as the smallest network to implement quantum computation, in which the highly entangled spin state, \textit{i.e.} the singlet, can be generated.
Requirements of precisely controlled qubits have intensively stimulated the detailed studies of the interactions in double-dot systems \cite{Hanson,Petta2} and the observations of phenomena thereby, such as Pauli spin blockade \cite{Petta2} and Coulomb blockade \cite{Coulomb1}. Furthermore, the demands for achieving efficient quantum computations and avoiding decoherence motivate us to manipulate spin states in double QDs in a fast and robust way. There are several methods to manipulate spin in QDs, such as electron spin resonance induced by magnetic field oscillating at the Zeeman transition frequency \cite{spin resonance} and electric control with spin-orbit (SO) coupling \cite{Rashba}. Recently, conventional ``rapid" adiabatic passages in quantum optics, for example, Landau-Zener scheme, have been extensively used to spin control in single QD \cite{RAP-single-dot}, coupled double QD \cite{RAP-DQD}, tripled QD \cite{RAP-tripled-dot},
which can be applied to prepare entanglement states \cite{QD-entanglement} and quantum logical gates, such as NOT \cite{Kestner} and CNOT \cite{Klinovaja} gates.

Shortcuts to adiabaticity \cite{Chen10a,Chen10b} have been proposed to speed up the adiabatic process without final excitation with many applications in atomic, molecular, optical physics, many-body physics, and even spintronics, see recent review \cite{review}. In a single QD, we applied the inverse engineering method \cite{ChenPRA} to design a fast and robust protocol of spin flip in the nanosecond timescale \cite{single-dot}, based on the Lewis-Riesenfeld invariant theory \cite{LR}. Furthermore, in a two-electron QD, more freedom in the applied electric fields provides the flexibility to control spin states by the invariant dynamics and controllable Lewis-Riesenfeld phases \cite{double-dot}. An alternative shortcut is provided by counter-diabatic control proposed by Demirplak and Rice \cite{Rice}, equivalent to tansitionless quantum driving \cite{Berry09}. This technique was
originally utilized to fast adiabatic control in two-level quantum systems theoretically \cite{Rice,Chen10b,Berry09} and experimentally \cite{Oliver,Suter}. Short afterwards, it has been extended to
multi-level systems \cite{Chen10b,2spin-Bmethod}, and even many-body systems \cite{Campo,Campo,Campo2,Campo3,Klaus}.

In this Report, we propose a fast and reliable protocol to generate the entangled spin states by using counter-diabatic driving. The external electric fields are designed for rapid spin control in a two-electron double QD in the presence of a static magnetic field and SO coupling. We apply the electric fields, instead of magnetic fields, and take advantage of SO coupling, since the time-dependent electric fields are easy to be generated on the nanoscale by adding local electrodes \cite{Nowack}. In addition, as comparing to a single QD, counter-diabatic driving is applicable in a two-electron double QD, as there exists more freedom with four controllable parameters, $x$ and $y$ components of the external electric fields for each dot. To simplify the experimental setup and reduce the device-dependent noise, we further apply the concept of multiple Schr\"{o}dinger pictures  \cite{Multiple-picture} to find an alternative shortcut with only $x$ component of the applied electric fields. Moreover, we also quantify how the electric fields
increase with shortening the time, to provide the lower bound of operation time for a given maximal amplitude of electric fields. Finally, the stability of designed shortcuts are discussed with respect to decoherence and systematic errors. Our approach presents a simple way to manipulate the singlet-triplet transition, which could be useful for rapid entanglement state preparation.

\section*{Results}
\begin{figure}[t]
\begin{center}
\scalebox{0.5}[0.5]{\includegraphics{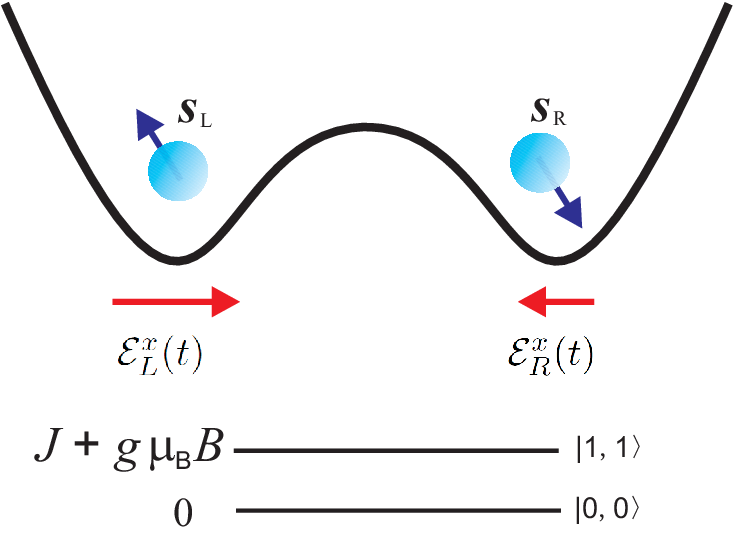}}
\caption{Schematic diagram of a two-electron double quantum dot in the presence of external electric fields and spin-orbit coupling,
where the singlet state and the lowest one of triplet states are considered as effective two-level system, when $|J+\Delta| \ll J$ with Zeeman term $\Delta =g \mu_B B$.}
\label{diagram}
\end{center}
\end{figure}

Two electrons are confined in a double QD, described as a quartic potential in Fig. \ref{diagram}, where they are isolated by Coulomb blockade \cite{Coulomb1}. In the presence of the applied magnetic fields, the lowest four eigenstates of the system can be expressed by singlet and triplet for $S=0$ and $S=1$ in the basis of $|S,S_z\ra$. This report presents a method to achieve fast adiabatic transition between the triplet and the singlet. We design the electric fields in $x-y$ plane to manipulate spin states with static magnetic fields along $z$ direction in each dot, considering structure-related Rashba ($\alpha$) and bulk-originated Dresselhaus ($\beta$) for [110] growth axis. If the energy difference between the singlet and the lowest one of the triplet is much less than the gap between the singlet and the triplet, we focus on the state transition between these lowest two, as shown in Fig. \ref{diagram}, where Land\'{e} factor $g < 0$ like in GaAs and InAs QDs.

%(counter-diabatic) interactions are designed to cancel the diabatic couplings of a reference process, thus accelerating the adiabatic process.
%%%%%%%%%%%%%%%%%%
%%%

By choosing $ |1 \ra = (1, 0)^T$ and $ |-1 \ra = (0, 1)^T$, referring to the states $|0,0 \ra$ and $|1,1 \ra$, respectively, we may first take the reference Hamiltonian as
\beqa
\label{H0}
H_0 = \frac{\hbar}{2} \left(
\begin{array}{cc}
Z & i Y
\\ - i Y & -Z
\end{array}
\right),
\eeqa
where
$
Y =-\sqrt{2}\alpha e (A^x_{L} - A^x_{R})/\hbar c
$,
$
Z = (-J - \Delta)/\hbar + e \beta (A^x_L + A^x_R) /\hbar c
$, and $A^x_j$ is determined by the electric fields, ${\bf \mathcal{E}}_j (t) = -(1/c)\partial \textbf{A}_j/\partial t$.
The subscriptions $j= L, R$ represent the left and the right dots, respectively. Here we assume the ansatz of the vector potentials is $A^x_{j} = A_0 \{\tanh[(t - a_j t_f)/(w_j t_f)] +1\}$, where $a_L=0.54$, $a_R=0.48$, $w_L=w_R=0.1$. The ansatz of vector potentials satisfies the condition $A_j^x(0)\simeq0$ and guarantees that the electric fields $\mathcal{E}^x_j$ start to be driven from $t=0$, that is, $\mathcal{E}^x_j \equiv 0$, when $t \leq 0$. 
%Meanwhile, the mixing angle $\theta$ goes from $\pi$ to $0$, crossing the point $\pi/2$ during the interval $(0,t_f)$.
When the adiabatic condition
\beqa
\label{condition}
\left|\frac{Z \dot{Y} - Y \dot{Z}}{(Y^2+Z^2)^{3/2}}\right| \ll 1
\eeqa
is fulfilled, the spin state will evolves from $|-1 \ra$ to $|1 \ra$ adiabatically along one of instantaneous eigenstates. When the final time is $t_f = 11$ ns,
the spin state is completely inverted, and the final population of $|1\ra$ is larger than $0.9999$.

\begin{figure}[t]
\begin{center}
\scalebox{0.5}[0.5]{\includegraphics{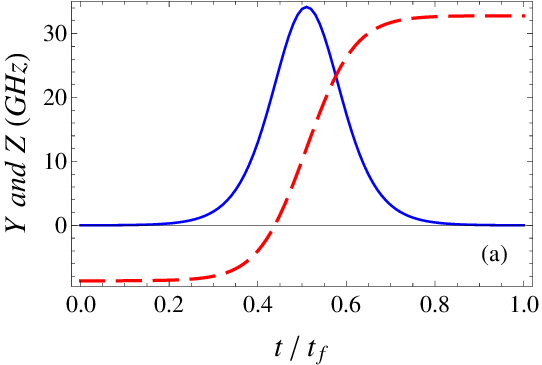}}
\scalebox{0.5}[0.52]{\includegraphics{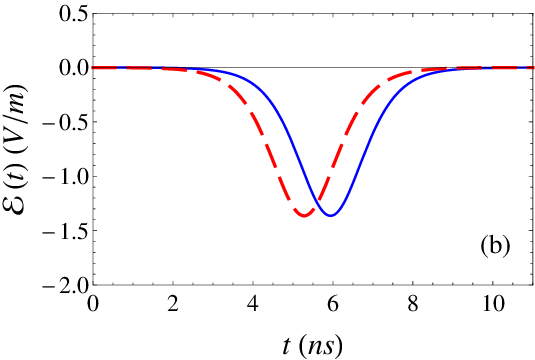}}
\scalebox{0.5}[0.52]{\includegraphics{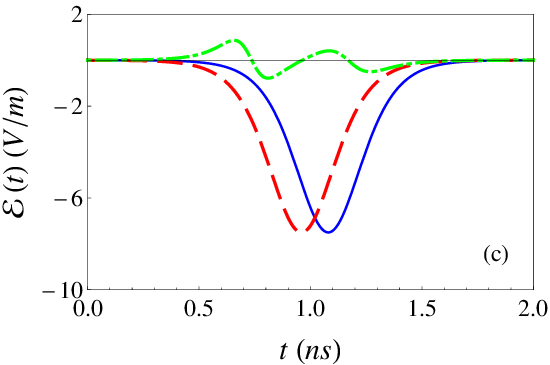}}
\caption{(a) Time dependence of $Y$ (solid blue line) and $Z$ (dashed red line) terms of $H_0$. (b) The applied electric fields ${\mathcal E}^x_L$ (solid blue line) and ${\mathcal E}^x_R$ (dashed red line) drive the state transition of $H_0$ adiabatically, with $t_f =11$ ns. (c) The applied electric fields ${\mathcal E}^x_L$ (solid blue line), ${\mathcal E}^x_R$ (dashed red line) and ${\mathcal E}^y_D$ (dot-dashed green line) drive the state transition of $H$ in a fast adiabatic way with shorter time $t_f=2$ ns.} \label{fig2}
\end{center}
\end{figure}

Shortening the manipulation time to $t_f = 2$ ns, shrinking $A_j^x$ into this time duration and keeping the same amplitude, we can find the state evolution is no longer adiabatic and the final state cannot reach $|1\ra$ at the final time. The same profiles of time-dependent $Y$ and $Z$ terms in $H_0$
are shown in Fig. \ref{fig2} (a) for different operation times, $t_f$.

%%%
%%%
Counter-diabatic driving, equivalent to transitionless quantum driving \cite{Berry09,Chen10b,Rice}, provides supplementary time-dependent interactions $H_1$ to cancel the diabatic couplings of $H_0$, and make the process fast and adiabatic, where $H_1$ is \cite{Chen10b}
\beqa
\label{H1}
H_1 = \frac{\hbar}{2} \left(
\begin{array}{cc}
0 & X
\\ X & 0
\end{array}
\right),
\eeqa
%, by means of adding two components of the electric fields in $y$ direction
with
$X = \sqrt{2}\alpha e (A^y_{L} - A^y_{R})/\hbar c$, driven by ${\mathcal E}^y_D$,
the difference between $y$ component of two electric fields. As a result, the exact dynamical evolution of total Hamiltonian $H=H_0+H_1$ coincides with adiabatic approximation of the reference Hamiltonian $H_0$. However, to implement accelerated adiabatic transitions more energy price has to pay, that is,
the maximal amplitude of $A^{y}_j$ in the $X$ term increases when the finally time $t_f$ is shortened.
This can be intuitively understood from time-energy uncertainty principle, that is, $A^{y}_j$ is proportional to $1/t_f$.
Since $ {\bf \mathcal{E}}_j (t) = -(1/c)\partial \textbf{A}_j/\partial t$, the larger value of ${\mathcal E}^x_j$ and ${\mathcal E}^y_D$ are finally required for the shorter time, $t_f$, as shown in Fig. \ref{fig2} (c).

In reality, the electron spin is subject to the device-dependent noise, which could be the amplitude noise of the electric fields \cite{single-dot}. It can be quite important, especially when the electric fields are relatively weak. From the above analysis, we find that four controllable parameters, ${\mathcal E}^x_{j}$ and ${\mathcal E}^y_{j}$, $x$ and $y$ components of the electric fields for each electron in a double QD should be applied. If $y$ component of the electric fields can be reduced, we can
remove the amplitude noise from $y$ component of the electric field. In addition to decreasing the total decoherent effects resulting from the device-dependent noise, the cancellation
of $y$ component of the electric field might be also useful to simplify the setup. To this end, we apply the concept of multiple Schr\"{o}dinger pictures to find an alternative way to implement the shortcuts. Making unitary transformation of Hamiltonian $H$ \cite{Multiple-picture,Berry90}
%%%
\begin{figure}[] \begin{center}
\scalebox{0.7}[0.7]{\includegraphics{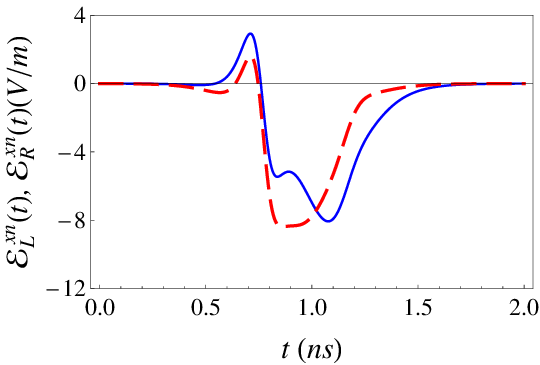}}
\caption{Electric fields of ${\mathcal E}^{xn}_L$ (solid blue line) and ${\mathcal E}^{xn}_R$ (dashed red line), designed from the Hamiltonian $\Tilde{H}$, see Eq. (\ref{Hs'}). } \label{fig3}
\end{center}
\end{figure}
by a rotation around $z$ axis with the angle $\pi/2-\phi$,
we obtain
\beqa
\label{Hs'}
\Tilde{H}   = \frac{\hbar}{2} \left(
\begin{array}{cc}
Z+\dot\phi & i Q
 \\ -i Q & -Z-\dot\phi
\end{array}
\right),
\eeqa
%%
%\beqa
%\label{H_I}
%H_I = \frac{\hbar}{2} \left(
%\begin{array}{cc}
%Z+\dot\phi & i Q
% \\ -i Q & -Z-\dot\phi
%\end{array}
%\right)
%\eeqa
%
without $\sigma_x$ term, where $\tan \phi = Y / X$ and $Q = \sqrt{X^2+Y^2}$. Again, the maximal amplitude of $Q$ will increase when decreasing time $t_f$, due to the
fact that $X$ becomes dominant (the maximal amplitude of $Y$ is unchanged).
%The dynamics in a common interaction picture may correspond to another Schr\"{o}dinger picture that represents a different physical process.
%Using $U'=1$, we may obtain
%
%
%where $K'= i \hbar \dot U' U'^\dag=0$.
The Hamiltonian $\Tilde{H}$ is equal to the original one $H$ at $t=0$ and $t_f$, which guarantees that the initial (final) states of $H$ and $\Tilde{H}$ coincide. However, the dynamics is not same during the intermediate process, although the populations are always equal.  Accordingly, we may acquire two new controllable $x$ component of the electric fields, $\mathcal{E}^{xn}_L$ and $\mathcal{E}^{xn}_R$, calculated from Eq. (\ref{Hs'}), see Fig. \ref{fig3}.

\section*{Discussions}

\begin{figure}[] \begin{center}
\scalebox{0.7}[0.7]{\includegraphics{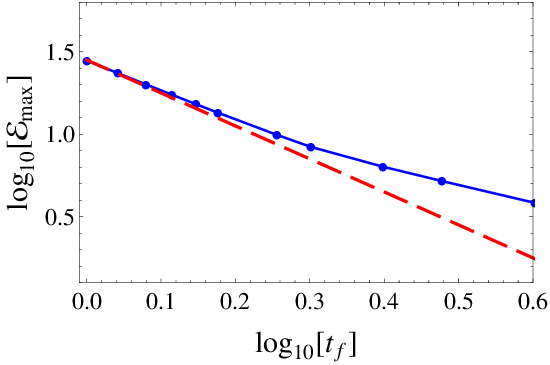}}
\caption{ Dependence of ${\mathcal E_{\textrm{max}}}$ on short time $t_f$ (solid blue line), where the dashed straight line shows
the asymptotic exponent of $t_f$, i.e. ${\mathcal E}_{\textrm{max}} \propto 1/t_f^2 $.
 } \label{fig4}
\end{center}
\end{figure}

Comparisons of $\mathcal{E}^{xn}_L$ and $\mathcal{E}^{xn}_R$ provided by different times suggest that stronger electric fields have to be used for shorter times, though
the amplitude of electric fields might be optimized by using superadiabatic iterations \cite{Multiple-picture}.
However, the amplitude of electric fields cannot be arbitrarily large simply because strong fields may destroy the systems. In order to quantify the energy price
mentioned above, we demonstrate the relation between the maximal values of electric fields and the operation time $t_f$, see Fig. \ref{fig4}. The maximal amplitude of
electric fields, ${\mathcal E}_{\max} = \max(|{\mathcal E}^{xn}_L|,|{\mathcal E}^{xn}_R|)$, fulfills the scaling law at very short times,
\beqa
\label{relation}
{\mathcal E}_{\max} \propto \frac{1}{t_f^2},
\eeqa
since $\mathcal{E}^{xn}_j \propto A_j^{y}/t_f$ and $A_j^{y} \propto 1/t_f$ go to infinity in the limit of $t_f \rightarrow 0$.
The asymptotic exponent of $t_f$ implies that the minimal time should be $\propto {\mathcal E}_{\max}^{-1/2}$, which provides the lower bound of
operation time when the maximal amplitude of electric fields is given. If the spin system in quantum dot, rather than the atom in harmonic trap,
is considered as working medium in the cooling cycles of quantum refrigerator, the minimal time for the (accelerated) adiabatic process, bounded by the energy, could be
relevant to the third law of thermodynamics and the unattainability principle \cite{Kosloff,energy-cost}.
%This is related to the third principle of thermodynamics and quantum speed limits.
%%%

\begin{figure}[]
\begin{center}
\scalebox{0.7}[0.7]{\includegraphics{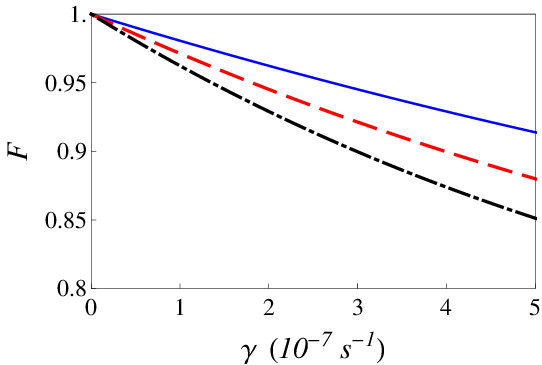}}
\caption{Fidelity $F$ versus dephasing rate $\gamma$ with respect to $t_f = 2$ ns (solid blue line), $t_f = 3$ ns (dashed red line), $t_f = 4$ ns (dot-dashed black line).}
\label{decoherence}
\end{center}
\end{figure}

For a realistic setup, the coupling to the stochastic environment is a general scenario to be considered,
where the hyperfine interactions with the nuclear spin could play important role at low temperature. To study the decoherence effect,
we present the master equation for the density matrix \cite{Sipe} in a generic form:
\begin{eqnarray}
\label{master equation}
\dot{\rho}  &=& -\frac{i}{\hbar} [\Tilde{H},\rho]- \frac{\gamma}{2} \sum_i [\sigma_i,[\sigma_i,\rho]]
\end{eqnarray}
where $\gamma$ is the dephasing rate. Solving the Bloch equation, we can obtain the final fidelity ($F=\rho_{11}$) for different times, see Fig. \ref{decoherence}, and
demonstrate that the faster manipulation increases the fidelity with less influences attributed by decoherence.
\begin{figure}[]
\begin{center}
\scalebox{0.7}[0.7]{\includegraphics{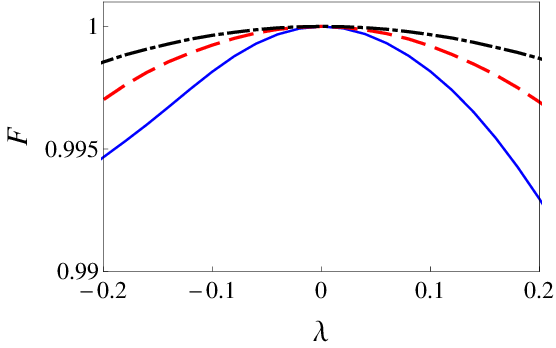}}
\caption{Fidelity $F$ versus $\lambda$ with respect to $t_f = 2$ ns (solid blue line), $t_f = 3$ ns (dashed red line), $t_f = 4$ ns (dot-dashed black line).}
\label{errors}
\end{center}
\end{figure}

To demonstrate the feasibility of our protocol, we also check the stability with respect to systematic errors in $\mathcal{E}^{xn}_j$.
The real electric fields can be $\mathcal{E}^{real}_j = \mathcal{E}^{xn}_j (1+ \lambda)$, where $\lambda$ is the relative deviation.
The dependence of fidelity $F$ on $\lambda$ is exhibited in Fig. \ref{errors} for different times.
Different from decoherence affected by the stochastic environment, fidelity is more stable with larger $t_f$, since the systematic error considered here depends on
the amplitude of electric fields. In general, the speeded-up adiabatic protocol has different stability with respect to different types of noise and systematic errors.
Alternatively, one can combine the inverse engineering and optimal control theory to pick up the most robust protocol in quantum two-level systems in presence of different noise and errors \cite{noise-NJP,Lu,Guerin}.

%This work not only provides an alternative shortcut to realize the fast adiabatic spin control but also proposes an efficient way to creat the entangled state by transitionless quantum driving. The Hamiltonian $H$ is composed by two parts, the reference process $H_0$, driven by $x$ component of electric fields, and the supplementary time-dependent interaction $H_1$, driven by $y$ component of electric fields. By applying $x$ and $y$ components of electric fields for each electron, the spin system follows exactly the adiabatic approximation of the reference Hamiltonian $H_0$, in the nanosecond timescale. In order to simplify the setup and decrease the device-dependent noise effect, we further transform the Hamiltonian by a unitary operator and obtain the new Hamiltonian implemented only by $x$ component of electric fields. To show the feasibility of our approach, we also check the fidelity with decoherence and systematic errors, respectively. We hope these results may lead to the applications in quantum information processing with the state-of-the-art technique.

\section*{Methods}
\textbf{Effective Hamiltonian.} The total spin-dependent Hamiltonian consists of Heisenberg term, Zeeman term, and interactions between the electric fields and the electrons, expressed as
%%%%%%%%%%%%%%%%%%
\beqa
\label{Hs}
H_{\textrm{total}} = J \bm{s}_L \cdot \bm{s}_R + \sum_j \Delta_j s^z_j - \frac{e}{c} \sum_j \textbf{A}_j \cdot \textbf{v}_j,
\eeqa
%%%%%%%%%%%%%%%%%%
The subscripts $j=L, R$ represent the left dot and the right one, respectively. Zeeman term is $\Delta =g \mu_B B$ with the equal magnetic fields $B$ applied to the left dot and the right one in $z$ direction, and $\textbf{A}_j$ are the vector potentials of the electric fields. The spin operators of two electrons confined in each dot are $\bm{s}_j = \bm{\sigma}_j/2$ with $z$ component $s^z_j$. The Heisenberg term $J \bm{s}_L \cdot \bm{s}_R$ describes the exchange coupling $J$ between two spins.
The example of a double QD of GaAs-based structure ($g=-0.44$) is taken with $B=3.7$ T. The energy gap between the singlet and the triplet is $J = 0.1$ meV, so that $|J + \Delta| / J = 0.06 \ll 1$. SO coupling term of Hamiltonian includes structure-related Rashba ($\alpha$) term and bulk-originated Dresselhaus ($\beta$) term for $[1 1 0]$ growth axis,
%%%%%
\begin{eqnarray}
\label{Hsoc}
H_{\rm{soc}} = \sum_j \alpha (\sigma^x_j p^y_j - \sigma^y_j p^x_j) + \sum_j \beta \sigma^z_j p^x_j,
\end{eqnarray}
%%%%%
so that the spin-dependent velocity operators become
%%%%%
\beqa
v^{x(y)}_j  &=& \frac{i}{\hbar} \left[H_{\rm soc}, x(y)_j\right].
\eeqa
%%%%%
As a result, after shifting some quantity of $H_{\rm{total}}$, we can derive a $2\times2$ Hamiltonian
%After shifting $H_{\textrm{total}}$ by some quantity, we derive the symmetric Hamiltonian,
%%%%%
\beqa
\label{H}
H = \frac{\hbar}{2} \left(
\begin{array}{cc}
Z & X + i Y
\\ X - i Y & -Z
\end{array}
\right),
\eeqa
%%%%%
where $Z$, $Y$ are $A_j^x$-dependent while $X$ is $A_j^y$-dependent, seen in the section above.

\textbf{Counter-diabatic driving and Z-axis rotation.}
Naturally, we separate the Hamiltonian $H$ into two parts, $H_0$ and $H_1$, where $H_0$ includes the $Y$ and $Z$ terms driven by
the $x$ components of electric fields applied in each dot, and $H_1$ includes only $X$ term driven by the $y$ components. The strategy of counter-diabatic driving
in a two-electron double QD is to set $H_0$ as reference first, which could be not adiabatic at all. Next,
we calculate and add the complementary interaction $H_1$ to cancel the diabatic coupling of
$H_0$ and make the spin control fast and adiabatic \cite{Chen10b,Rice,Berry09}. Actually, the separation of Hamilton $H$ (\ref{H}) into $H_0$ and $H_1$ depends strongly
on the choice of growth axis [110]. For instance, if the growth axis [111] is chosen, the SO coupling term should be modified as
\beq
H_{\rm{soc}} = \sum_j \alpha (\sigma^x_j p^y_j - \sigma^y_j p^x_j) + \sum_j \beta \sigma^y_j p^x_j,
\eeq
and the $2\times2$ Hamiltonian (\ref{H}) becomes
\beqa
H = \frac{\hbar}{2} \left(
\begin{array}{cc}
-(J+\Delta)/\hbar & X + i Y
\\ X - i Y & (J+\Delta)/\hbar
\end{array}
\right),
\eeqa
with $X = \sqrt{2}\alpha e (A^y_{L} - A^y_{R})/\hbar c$ and $Y =-\sqrt{2} (\alpha+\beta) e (A^x_{L} - A^x_{R})/\hbar c$.
Therefore, the approach presented here is not valid, since the reference $H_0$ and the counter-diabatic driving $H_1$ can not be naturally separated and calculated.

Here counter-diabatic driving is applicable in a two-electron double QD, as in Hamiltonian $H$ (\ref{H}) there exists freedom with four controllable parameters, $x$ and $y$ components of the external electric fields for each dot. This is different from the Hamiltonian in a single QD where there are only two controllable parameters, $x$ and $y$ components of the electric field, so that it is impossible to produce the required all-electrical interaction by counter-diabatic driving \cite{single-dot}.

%\textbf{Master equation}
%The master equation for the density matrix is presented as Eq. \ref{master equation}. We define the Bloch vector with components $u = \rho_{1-1} + \rho_{-11}$, $v = -i(\rho_{1-1} - %\rho_{-11})$ and $w = \rho_{11} - \rho_{-1-1}$ and thus obtain the Bloch equation,
%
%\beqa
%\label{bloch equation}
%\left(\begin{array}{ccc}
% \dot{u}
%\\
%\dot{v}
%\\
%\dot{w}
%\end{array}\right)
%=
%\left(\begin{array}{ccc}
%-4 \gamma & Z+\dot{\phi} & - Q
%\\
%-Z-\dot{\phi} & -4 \gamma & 0
%\\
%Q  & 0 & -4 \gamma
%\end{array}\right)
%\left(\begin{array}{ccc}
%u
%\\
%v
%\\
%w
%\end{array}\right).
%\eeqa
%
Furthermore, using multiple Schr\"{o}dinger pictures to describe various physical settings sharing the same dynamics is
helpful to find alternative shortcuts, when the counter-diabatic term is difficult or impossible to implement \cite{Multiple-picture}.
One can transform the Hamiltonian  based on Lie algebra to cancel the unwanted component of Hamiltonian \cite{Muga}.
Applying this concept, we make unitary transformation of Hamiltonian $H$ by z-axis rotation. While the original dynamics satisfies $i \hbar \partial_t \Psi(t) = H \Psi(t)$, the new dynamics is given by $i \hbar \partial_t \Tilde{\Psi}(t) =  \Tilde{H} \Tilde{\Psi}(t)$, where $\Tilde{\Psi}(t) = U^\dag \Psi(t)$, $\Tilde{H} = U^\dag (H-K) U$ and $K=i\hbar\dot{U} U^\dag$.

%\section*{References}

\section*{Acknowledgement}
\noindent We appreciate E. Ya Sherman for his fruitful discussions. This work is partially supported by the NSFC (61176118), the Shanghai Rising-Star, Pujiang and Yangfan Program (12QH1400800, 13PJ1403000, 14YF1408400), the Specialized Research Fund for the Doctoral Program of Higher Education (2013310811003), and the Program for Professor of Special Appointment (Eastern Scholar) at Shanghai Institutions of Higher Learning.

%\section*{Author contributions}
%\noindent YB carried out the theoretical and numerical calculation; XC
%analyzed the theoretical results. Both authors wrote and reviewed the manuscript.

%\section*{Additional information}
%\noindent Competing financial interests: The authors declare no competing financial interests.


\begin{thebibliography}{10}

\bibitem{spin resonance} F. H. L. Koppens et al., ``Driven coherent oscillations of a single electron spin in a quantum dot", \textit{Nature} \textbf{442}, 766 (2006).

\bibitem{Rashba} E. I. Rashba, ``Theory of electric dipole spin resonance in quantum dots: Mean field theory with Gaussian fluctuations and beyond", \textit{Phys. Rev. B} \textbf{78}, 195302 (2008).%;Rashba E I and Efros Al L 2003 \textit{Phys. Rev. Lett.} \textbf{91} 126405

\bibitem{Nowack} K. C. Nowack, F. H. L. Koppens, Y. V. Nazarov, and L. M. K. Vandersypen, ``Coherent control of a single electron spin with electric fields", \textit{Science} \textbf{318}, 1430 (2007).

\bibitem{You-Sun} R. Li, J. Q. You, C. P. Sun, and F. Nori, ``Controlling a nanowire spin-orbit qubit via electric-dipole spin resonance", \textit{Phys. Rev. Lett.} \textbf{111}, 086805 (2013).

\bibitem{Guo} G. Cao et al., ``Ultrafast universal quantum control of a quantum-dot charge qubit using Landau-Zener-St\"{u}ckelberg interference", \textit{Nat. Comm.} \textbf{4}, 1401 (2013).

\bibitem{QD-Qgates} G. Burkard, D. Loss, and D. P. DiVincenzo, ``Coupled quantum dots as quantum gates", \textit{Phy. Rev. B} \textbf{59}, 2070 (1999).

\bibitem{Hanson} R. Hanson, L. P. Kouwenhoven, J. R. Petta, S. Tarucha, and L. M. K. Vandersypen, ``Spins in few-electron quantum dots", \textit{Rev. Mod. Phys.} \textbf{79}, 1217 (2007).

\bibitem{Petta2} J. M. Taylor, J. R. Petta, A. C. Johnson, A. Yacoby, C. M. Marcus, and M. D. Lukin, ``Relaxation, dephasing, and quantum control of electron spins in double quantum dots", \textit{Phys. Rev. B} \textbf{76}, 035315 (2007).

\bibitem{Coulomb1} F. R. Waugh, M. J. Berry, D. J. Mar, R. M. Westervelt, K. L. Campman, and A. C. Gossard, ``Single-electron charging in double and triple quantum dots with tunable coupling", \textit{Phys. Rev. Lett.} \textbf{75}, 705 (1995); C. Livermore, C. H. Crouch, R. M. Westervelt, K. L. Campman, and A. C. Gossard, ``The Coulomb blockade in coupled quantum dots", \textit{Science} \textbf{274}, 1332 (1996).


\bibitem{RAP-single-dot} M. Shafiei, K. C. Nowack, C. Reich, W. Wegscheider, and L. M. K. Vandersypen, ``Resolving spin-orbit- and hyperfine-mediated electric dipole spin resonance in a quantum dot", \textit{Phys. Rev. Lett.} \textbf{110}, 107601 (2013).

\bibitem{RAP-DQD} H. Ribeiro, G. Burkard, J. R. Petta, H. Lu, and A. C. Gossard, ``Coherent adiabatic spin control in the presence of charge noise using tailored pulses", \textit{ Phys. Rev. Lett.} \textbf{110}, 086804 (2013).

\bibitem{RAP-tripled-dot} J. Huneke, G. Platero, and S. Kohler, ``Steady-state coherent transfer by adiabatic passage", \textit{Phys. Rev. Lett. } \textbf{110}, 036802 (2013).

\bibitem{QD-entanglement} C. Creatore, R. T. Brierley, R. T. Phillips, P. B. Littlewood, and P. R. Eastham, ``Creation of entangled states in coupled quantum dots via adiabatic rapid passag", \textit{Phys. Rev. B} \textbf{86}, 155442 (2012).

\bibitem{Kestner} J. P. Kestner and S. D. Sarma, ``Proposed spin-qubit controlled-not gate robust against noisy coupling", \textit{Phy. Rev. A} \textbf{84}, 012315 (2011).

\bibitem{Klinovaja} J. Klinovaja, D. Stepanenko, B. I. Halperin, and D. Loss, ``Exchange-based CNOT gates for singlet-triplet qubits with spin-orbit interaction", \textit{ Phys. Rev. B} \textbf{86}, 085423 (2012).

\bibitem{Chen10a} X. Chen, A. Ruschhaupt, S. Schmidt, A. del Campo, D. Gu\'{e}ry-Odelin, and J. G. Muga, ``Fast optimal frictionless atom cooling in harmonic traps: Shortcut to adiabaticity", \textit{Phys. Rev. Lett.} \textbf{104}, 063002 (2010).

\bibitem{Chen10b} X. Chen, I. Lizuain, A. Ruschhaupt, D. Gu\'{e}ry-Odelin, and J. G. Muga, ``Shortcut to adiabatic passage in two-and three-level atoms", \textit{ Phys. Rev. Lett.} \textbf{105}, 123003 (2010).

\bibitem{review} E. Torrontegui et al., ``Shortcuts to adiabaticity", \textit{Adv. At. Mol. Opt. Phys.} \textbf{62}, 117 (2013).


\bibitem{ChenPRA} X. Chen, E. Torrontegui, and J. G. Muga, ``Lewis-Riesenfeld invariants and transitionless quantum driving", \textit{Phys. Rev. A} \textbf{83}, 062116 (2011).

\bibitem{single-dot} Y. Ban, X. Chen, E. Ya Sherman, and J. G. Muga, ``Fast and robust spin manipulation in a quantum dot by electric fields",
    \textit{Phys. Rev. Lett.} \textbf{109}, 206602 (2012).

\bibitem{LR} H. R. Lewis and W. B. Riesenfeld, ``An exact quantum theory of the time-dependent harmonic oscillator and of a charged particle in a time-dependent electromagnetic field", \textit{J. Math. Phys.} \textbf{10}, 1458 (1969).


\bibitem{double-dot} Y. Ban, X. Chen, J. G. Muga, and E. Ya. Sherman, ``Fast spin control in a two-electron double quantum dot by dynamical invariants", arXiv:1309.1916.

\bibitem{Rice} M. Demirplak and S. A. Rice, ``Adiabatic population transfer with control fields", \textit{J. Phys. Chem. A} \textbf{107}, 9937 (2003); ``Assisted adiabatic passage revisited", \textit{J. Phys. Chem. B} \textbf{109}, 6838 (2005); ``On the consistency, extremal, and global properties of counterdiabatic fields", \textit{J. Chem. Phys.} \textbf{129}, 154111 (2008).

\bibitem{Berry09} M. V. Berry, ``Transitionless quantum driving", \textit{J. Phys. A} \textbf{42}, 365303 (2009).

\bibitem{Oliver} M. G. Bason et al., ``High-fidelity quantum driving", \textit{Nat. Phys.} \textbf{8}, 147 (2012).

\bibitem{Suter} J.-F. Zhang et al., ``Experimental implementation of assisted quantum adiabatic passage in a single spin", \textit{Phys. Rev. Lett.} \textbf{110}, 240501 (2013).
%Experimental implementation of assisted quantum adiabatic passage in a single spin


\bibitem{2spin-Bmethod} K. Takahashi, ``Transitionless quantum driving for spin systems", \textit{Phys. Rev. E} \textbf{87}, 062117 (2013).

\bibitem{Klaus} T. Opatrn\'{y} and K. M{\o}lmer,  ``Partial suppression of nonadiabatic transitions", \textit{New J. Phys.} \textbf{16}, 015025 (2014).

\bibitem{Campo} A. del Campo, M. M. Rams, and W. H. Zurek, ``Assisted finite-rate adiabatic passage across a quantum critical point: Exact solution for the quantum Ising model", \textit{Phys. Rev. Lett.} \textbf{109}, 115703 (2012).

\bibitem{Campo2} A. del Campo, ``Shortcuts to adiabaticity by counterdiabatic driving", \textit{Phys. Rev. Lett.} \textbf{111}, 100502 (2013).


\bibitem{Campo3} S. Deffner, C. Jarzynski, and A. del Campo, ``Classical and quantum shortcuts to adiabaticity for scale-invariant driving",
   \textit{Phys. Rev. X} \textbf{4}, 021013 (2014).


\bibitem{Multiple-picture} S. Iba\~{n}ez, X. Chen, E. Torrontegui, J. G. Muga, and A. Ruschhaupt, ``Multiple Schr\"{o}dinger pictures and dynamics in shortcuts to adiabaticity", \textit{Phys. Rev. Lett.} \textbf{109} 100403 (2012).

\bibitem{Berry90} M. V. Berry, ``Histories of adiabatic quantum transitions", \textit{Proc. R. Soc. A} \textbf{429}, 61 (1990).

\bibitem{Kosloff} Y. Rezek, P. Salamon, K. H. Hoffmann, and R. Kosloff, ``The quantum refrigerator: The quest for absolute zero", \textit{EPL} \textbf{85}, 60015 (2009).

\bibitem{energy-cost} X. Chen and J. G. Muga, ``Transient energy excitation in shortcuts to adiabaticity for the time dependent harmonic oscillator", \textit{Phys. Rev. A} \textbf{82}, 053403 (2010).

\bibitem{Sipe} K. S. Virk and J. E. Sipe, ``Conduction electrons and the decoherence of impurity-bound electrons in a semiconductor", \textit{ Phys. Rev. B} \textbf{72}, 155312 (2005).

\bibitem{noise-NJP} A. Ruschhaupt, X. Chen, D. Alonso, and J. G. Muga, ``Optimally robust shortcuts to population inversion in two-level quantum systems", \textit{New J. Phys.} \textbf{14}, 093040 (2012).

\bibitem{Lu} X.-J. Lu, X. Chen, A. Ruschhaupt, D. Alonso, S. Gu\'{e}rin, and J. G. Muga, ``Fast and robust population transfer in two-level quantum systems with dephasing noise and/or systematic frequency errors", \textit{Phys. Rev. A} \textbf{88}, 033406 (2013).

\bibitem{Guerin} D. Daems, A. Ruschhaupt, D. Sugny, and S. Gu\'{e}rin, ``Robust quantum control by a single-shot shaped pulse", \textit{Phys. Rev. Lett.} \textbf{111}, 050404 (2013).

%\bibitem{optimization2} G. Dridi, S. GušŠrin, V. Hakobyan, H. R. Jauslin, and H. Eleuch, Phys. Rev. A \textbf{80}, 043408 (2013).
%\bibitem{optimization3} S. Gu\'{e}rin, V. Hakobyan, and H. R Jauslin, Phys. Rev. A \textbf{80}, 043408 (2013).

\bibitem{Muga} E. Torrontegui, S. Mart\'{\i}nez-Garaot, and J. G. Muga, ``Hamiltonian engineering via invariants and dynamical algebra", \textit{Phys. Rev. A} \textbf{89}, 043408 (2014).

\end{thebibliography}
\end{document}